# Probing the Local Dielectric Function of $WS_2$ on an Au Substrate by Near Field Optical Microscopy Operating in the Visible Spectral Range

Authors:

Oisín Garrity[a], Alvaro Rodriguez[b], Niclas S. Mueller[c], Otakar Frank[b], Patryk Kusch[a]*

Affiliations:

a) Department of Physics, Freie Universität Berlin, Arnimallee 14, D-14195 Berlin, Germany
b) J Heyrovský Institute of Physical Chemistry, Academy of Sciences of the Czech Republic, Dolejškova 3, CZ-18223 Prague 8, Czechia
c) NanoPhotonics Centre, University of Cambridge, UK

*corresponding author: patryk.kusch@fu-berlin.de

## Abstract

The optoelectronic properties of nanoscale systems such as carbon nanotubes (CNTs), graphene nanoribbons and transition metal dichalcogenides (TMDCs) are determined by their dielectric function. This complex, frequency dependent function is affected by excitonic resonances, charge transfer effects, doping, sample stress and strain, and surface roughness. Knowledge of the dielectric function grants access to a material's transmissive and absorptive characteristics. Here we use the dual scanning near field optical microscope (dual s-SNOM) for imaging local dielectric variations and extracting dielectric function values using a pre-established mathematical inversion method. To demonstrate our approach, we studied a monolayer of $WS_2$ on bulk Au and identified two areas with differing levels of charge transfer. The experiments highlight a further advantage of the technique: the dielectric function of contaminated samples can be measured, as dirty areas can be easily identified and excluded for the calculation, being important especially for exfoliated 2D materials [1]. Our measurements are corroborated by atomic force microscopy (AFM), Kelvin force probe microscopy (KPFM), photoluminescence (PL) intensity mapping, and tip enhanced photoluminescence (TEPL). We extracted local dielectric variations from s-SNOM images and confirmed the reliability of the obtained values with spectroscopic imaging ellipsometry (SIE) measurements.

## 1. Introduction

Advanced photonic and optoelectronic based integrated circuits (ICs) and devices depend on novel nanoscale materials such as CNTs, graphene nanostructures such as nanoribbons, and 2D materials like TMDCs. CNTs have already shown promise in creating novel optoelectronic ICs [2], [3], while graphene nanoribbons have been used in novel photonic ICs [4]. TMDCs have favourable properties for photonic applications due to excitonic states that exist at room temperature and the direct bandgap in monolayers, which have been used to realise photodetectors on photonic ICs [5]. To advance the applications of these nanoscale materials in photonics it is essential to characterise their fundamental electrical and optical properties like, for example, transmittance and absorption in addition to their frequency dependence.

An elegant way to gain access to the electric and optical characteristics of nanoscale materials is through determining their complex dielectric function which is a measure of the transmission and absorption of light through a material as a function of frequency [6]. This frequency dependence is not linear as it is affected by resonances around the exciton transition energies, a fact that enables the study of, for example, the influence of excitons on the optical properties of TMDC samples [6]. The dielectric function also provides insights into electrical and magnetic properties of nanoscale samples, important characteristics to consider when thinking about implementing them into novel applications.

The dielectric function is largely affected by material disorder. Extrinsic material disorder stems from the environment, while intrinsic disorder describes material disorder stemming from crystalline imperfections [7]. These effects can to some extent be characterised with established characterisation methods such as AFM and scanning tunnelling microscopy (STM) [7] in addition to far field methods like Raman and PL spectroscopies for identification of stress and strain [8] and KPFM for charge distribution information [9].

Along with these common sources of disorder, nanoscale systems add additional complexity stemming from their reduced dimensionality which results in a decrease of dielectric screening and an increase of the Coulomb interaction between charge carriers [10]. Considering TMDCs as an example, this has an effect on their electronic response such as allowing excitonic pairs to exist at room temperature as well as changes in the band gap when thinning TMDCs from multilayers to monolayers [11]. Inevitably intrinsic disorder like surface roughness, impurities, and defects will lead to a local fluctuation of the exciton binding energy as well as fluctuations in the bandgap leading to a new form of disorder now being referred to as dielectric disorder [10]. This form of disorder can affect samples down to the nanometre scale and can strongly affect their optical and electronic properties. To characterise this nanoscale form of disorder requires a technique with a spatial resolution on the nanometre scale that is sensitive to the local dielectric properties of the sample, and for 1D and 2D materials it needs to be surface sensitive.

The far field methods that are conventionally used to determine the dielectric function, like ellipsometry and reflectance experiments, can identify variations stemming from material disorder with a resolution down to the micrometre length scale [12]. Sample disorder and bandgap variations can be observed via PL/Raman spectroscopy [13]. However, the spatial resolution of these techniques would not serve well to investigate dielectric disorder due to the resolution being bound by the diffraction limit, which prevents nanometre resolution. AFM and KPFM can achieve nanometre resolution, and are used to image defects, intrinsic disorder, and surface potential information [14], [15]. SNOM itself was demonstrated recently in probing the dielectric screening of hBN in graphene integrated on silicon photonics at nanometre resolution [16]. Tip-enhanced Raman and PL spectroscopy can be implemented to identify nanoscale bandgap variations, strain, and defects [17], [18]. These techniques however are not suitable to quantify, or to identify nanoscale variations in, the dielectric function.

Here we show the dual s-SNOM as an advanced tool for measuring the dielectric function with nanometre scale resolution. To demonstrate this capability, we nano-imaged a sample consisting of monolayer tungsten disulphide ($WS_2$) placed on a gold (Au) substrate with the aim to identify local variations in the optical properties of the sample. We find that the intrinsic disorder such as surface roughness, boundaries, and charge transfer influence the dielectric function and can lead to strong near field contrast changes in the s-SNOM images of the $WS_2$. We recorded s-SNOM images at different harmonics and three different excitation wavelengths. We determined the dielectric values by using a pre-established inversion method that extracts the dielectric value from s-SNOM data (see Govyadinov et al. [19] and Tranca et al. [20]), as a function of tip position. The average dielectric values determined using the more volume sensitive second harmonic s-SNOM image data, gives the joint dielectric values of Au and $WS_2$ for three excitation frequencies that are in excellent agreement with ellipsometry measurements. Average dielectric values from the more surface sensitive fourth harmonic, in contrast, differ from the bulk sensitive ellipsometry measurements due to their different penetration depth. We demonstrate the resolution of the dual s-SNOM by comparing the higher harmonic near field changes with the TEPL peak position, showing an inverse correlation. We correlate the s-SNOM measurements with spatially resolved PL spectroscopy and KPFM measurements, where we observe areas of strongly quenched PL and lower surface potential. This indicates that the strong variations of the dielectric function in monolayer $WS_2$ are mainly due to charge transfer effects.

## 2. Material and Methods

The sample used in this work was made using a dry transfer method that utilises a viscoelastic polydimethylsiloxane (PDMS) stamp to mechanically exfoliate $WS_2$ crystals (HQ graphene) onto a $SiO_2$ substrate that had a 40nm layer of Au deposited on top using magnetron sputtering [21]. The approximate time between Au deposition and $WS_2$ stamping was 30 mins.

For the pre-characterisation of the sample, we used PL spectroscopy and KPFM measurements. The PL spectra were recorded with a Horiba Jobin-Yvon XploRA micro-Raman spectrometer as a function of laser position on the sample with a 0.90NA (NA: numerical aperature) 100x objective. For excitation we used a laser with 532nm wavelength and 1 mW power on the sample, with an acquisition time of 1 s and a 600 grooves per mm grating. KPFM images were obtained with an AIST-NT scanning probe microscope, using Pt-Ir coated Si tips (ACCESS-EFM probes, AppNano, k = 2.7 N.m$^{-1}$), and 1300 nm diode for the detection measurement. KPFM was operated in the amplitude modulated mode (AM-KPFM), which is sensitive to the electrostatic forces [22]. SIE was performed using an EP4 ellipsometer (Accurion Gmbh, Germany). Monochromatic light was provided by a xenon lamp with several interference filters. The angle of incidence was varied between 55 and 65°. The obtained data was analysed using Accurion's EP4Model software. Due to our sample being a monolayer, SIE is insensitive to the out-of-plane component so the dielectric function determined by SIE in this manuscript should be regarded as pseudo-dielectric.

The sample was characterised using dual s-SNOM and TEPL utilising a commercial s-SNOM (NeaSNOM from Neaspec GmbH, Germany). We used platinum-iridium coated AFM tips (23nm coating thickness) from Nano world, featuring a tip apex radius of below 25nm. A wavelength-tuneable cw laser (Hübner C-Wave, 450-650 nm wavelength) was used for excitation which was guided through a beam expander onto a parabolic mirror with a NA of 0.4. The parabolic mirror focusses the laser light onto the AFM tip which then acts as a near field probe in the visible spectral range. The parabolic mirror also collects the backscattered light. The laser power for all s-SNOM image measurements and TEPL measurements was ~1 mW at the tip with an integration time of 16 ms. The tip amplitude was 53.5 nm with a tapping frequency of 243 kHz. The tip-enhanced PL spectroscopy was taken via s-SNOM using side illumination equipped with a Kymera 328i spectrometer (Andor) and an Si CCD with a tip scanning step count of 50 nm per pixel. A sketch of the setup can be found in Ref[23].

Background suppression was done by oscillating the tip at an amplitude of 50nm at a frequency $\Omega$ of about 250 kHz. The signal was then demodulated at higher harmonics of the tip frequency $n\Omega$. The noise was further reduced by a pseudo-heterodyne interferometer which has a reference mirror that oscillates at frequency $M \ll \Omega$, changing the length of the reference beam path producing interference with the scattered signal. This produces sidebands around the fundamental harmonics at frequencies $f = n\Omega \pm mM$ where $m$ is an integer ≥ 1. Using this detection regime, the amplitude and the phase are recorded from the sample. Both are recorded as near field amplitude and phase images at various sidebands *m* of the fundamental harmonic. An increase in m leads to a decrease in noise present in the s-SNOM images.

## 3. Theory

To extract the dielectric constant of the sample at a particular wavelength from s-SNOM images, we follow an inversion method that was introduced in Ref. [19]. It is based on the prevailing theoretical description of the s-SNOM which describes the detected signal, containing both amplitude and phase information, by the scattering coefficient $\sigma = E_s/E_i$, where $E_s$ describes the electric field amplitude of the scattered light and $E_i$ describes the amplitude of the incident radiation [19], [24]. The electric field amplitude at the tip is given by $(1 + r_s)E_i$, which accounts for light that is reflected by the sample to the tip, with $r_s$ the reflection coefficient of the sample. This electric field polarises the tip, yielding an effective dipole $p = \alpha_{eff}(1 + r_s)E_i$, where $\alpha_{eff}$ is the effective polarizability of the tip that accounts for the near field interaction between sample and tip. The scattering of this effective dipole leads to an electric field amplitude $E_s \propto (1 + r_s)p$ at the detector, assuming that part of the scattered light is also reflected by the sample. Finally, the scattering coefficient can now be written as

$$\sigma \propto \alpha_{eff}(1 + r_s)^2 \qquad (1)$$

which is complex as $E_i$ and $E_s$ may have a phase difference [19], [24]. Thus, the signal that is recorded by the detector is proportional to the effective tip polarizability, which is determined by the interaction of the tip near field with the sample (see below).

The dual s-SNOM relies on this strongly enhanced and localised electromagnetic near field between the sample and the metallic tip which stems from mechanisms such as the lightning rod effect and localised surface plasmon resonances in the metallic tip [23]. To model this, we used a point dipole model which regards the tip apex as a perfectly conducting sphere with radius $R_t \ll \lambda$ [19][24]. By considering the sphere as a point dipole within the quasistatic approximation, one obtains an expression for the near field between the tip dipole and sample by considering a *mirror* point dipole whose direction is parallel to the tip dipole [24]. Solving these equations electrostatically yields the effective polarizability of the SNOM tip

$$\alpha_{eff} = \alpha_0[1 - f(H)\beta(\epsilon_s)]^{-1} \qquad (2)$$

which depends on two functions:

$$f(H) = \alpha_0/(16\pi(R_t + d_0 + H)^3) \qquad (3)$$

is related to the height $H$ of the tip above the sample, and $d_0$ the height above the Au substrate, with

$$\alpha_0 = 4\pi R_t^3(\epsilon_t - 1)/(\epsilon_t + 2) \qquad (4)$$

where $\epsilon_t$ is the dielectric function of the tip material, and

$$\beta(\epsilon_s) = (\epsilon_s - 1)/(\epsilon_s + 1) \qquad (5)$$

only depends on the dielectric function of the sample $\epsilon_s$ [24].

The dual s-SNOM demodulates the detected signal using a lock-in amplifier as well as a pseudo-heterodyne detection method [24] to suppress background. As a result of this background suppression, the dual s-SNOM generates near field images at different harmonics, with each harmonic being sensitive to different sample depths [25].

The demodulated detected signal at the $n^{th}$ harmonic described by a complex Fourier transform

$$\sigma_n(H) = \hat{F}_n[\sigma(\beta, H(t))] = \int \sigma(\beta, H(t)) \, exp(in\Omega t) \tag{6}$$

of the scattering coefficient $\sigma(\beta, H(t))$ from the tip. Since the scattering coefficient $\sigma$ depends on many unknown details of the detection pathway, we measure the near field contrast

$$\eta_n = \frac{\sigma_n}{\sigma_{n,ref}},$$

where $\sigma_{n,ref}$ is the scattering recorded at a reference sample position with known dielectric value $\epsilon_{s,ref}$ (in our case a bare Au substrate). For an unknown dielectric value of the sample $\epsilon_s$ it is non-trivial to calculate $\epsilon_s$ from $\eta_n$ because of the non-algebraic relation of $\sigma_n$ and $\epsilon_s$. We therefore use a Taylor expansion of the near field contrast (see Ref. [19] for details)

$$\eta_n = \sum_{j=1}^{J} \frac{\hat{F}_n[\alpha_0^j f^j]}{\hat{F}_n[\alpha_{eff,ref}]} \beta^j \tag{6}$$

where $\alpha_{eff,ref}$ is the tip polarizability at the reference sample and $\alpha_0^j$ an expansion coefficient. When truncating the Taylor expansion at a specific order $J$, Eq. (6) can be inverted to find $\beta$. Once $\beta$ is found, it is then trivial to recover the dielectric constant of the sample at a specific wavelength using the relation $\epsilon_s = (1 + \beta)/(1 - \beta)$ [19]. Since we know the distance $H$ between the sample and the tip, and we can input the tip-Au distance $d_0$ by extracting it from the AFM topography, we can use this inversion method to recover the spatially resolved $\epsilon_s$ from the s-SNOM images. By recording the s-SNOM images at different excitation energies $\omega$, this allows us to obtain the dispersion of the sample $\epsilon_s(\omega)$.

## 4. Results and Discussion

The WS$_2$/Au sample (Figure S1, Supplementary Information) was pre-characterised using KPFM and PL spectroscopy; KPFM provides surface potential information while at the same time providing an AFM image outlining the surface morphology of the sample. From the AFM image we identify a WS$_2$ monolayer that covers the Au substrate, Figure 1(a). The WS$_2$ is folded as a bilayer close to a wrinkle on the Au substrate identified as the magenta-shaded area. We furthermore identify an area of the WS$_2$ monolayer surrounding the bilayer that has a different contrast, identified by the blue-shaded region. For clarity of sample orientation for Figure 1(a), (b) and (c), a wrinkle has been highlighted as a red-dashed line.

Concerning the KPFM measurements (Figure 1(b) – same sample area as Figure 1(a), rotated by roughly 8.5 degrees clockwise), the contrast difference show areas of differing work functions, determined from the differences in contact potential between the sample and the tip [26]. We find that the surface potential of the WS$_2$ (blue-shaded area) is almost the same as of the Au substrate (yellow-shaded area) (145 mV) while the surrounding area shows a darker contrast (average surface potential of -175mV) indicating that charge transfer has occurred between the two materials. The work function of Au at room temperature is ~5.30 eV [27] while the work function of monolayer WS$_2$ is ~4.90 eV [28] indicating that charge transfer of electrons should take place from WS$_2$ into Au. We note that the KPFM measurements were performed in "dark" in relation to the possible exciton generation in WS2, because the cantilever deflection measurement is done using a 1300 nm laser diode. Hence, no photogenerated carriers are influencing the surface potential [29].

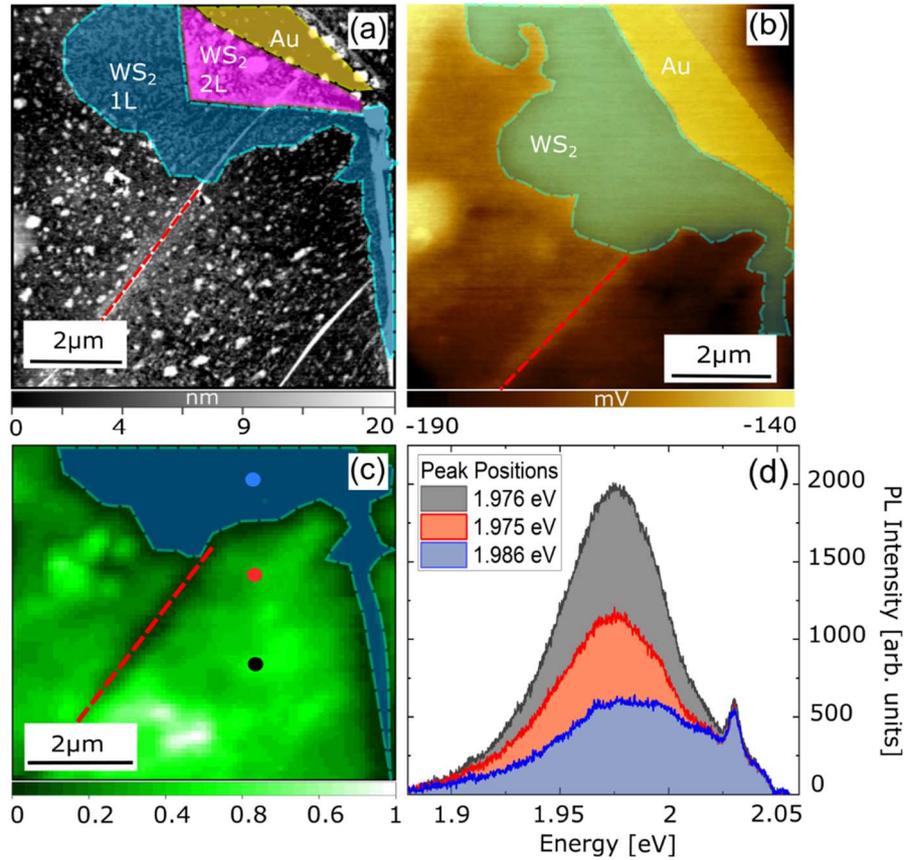

Figure 1: Sample pre-characterisation: (a) AFM topography (colour bar in nm), (b) KPFM (colour bar in mV). (d) TEPL spectra taken at respective coloured points indicated in (c) the PL intensity map. The blue shading shows a spectrum that was recorded at the areas where the WS$_2$ is more adhering to the Au substrate.

To gain more insight into the reason behind these areas of differing surface potential contrast, we conducted spatially resolved PL measurements [30]. By measuring the PL intensity as a function of laser position, we obtained a PL map, with a spatial resolution of ~570 nm [31] from a diffraction limited laser spot. Figure 1(c) shows the PL intensity map taken at the same sample area as the KPFM. The top part of the map (shaded in blue) shows an area of partially quenched PL, the borders of which bear a strong resemblance to the adhered area around the bilayer in the AFM and KPFM images. Quenched PL arises because of the charge transfer that has occurred in this area [32], backing up the KPFM findings. Moving beyond the borders of this adhered area, the PL increases in intensity while also exhibiting sizable peak shifts throughout this PL intense area of the sample (example shown in S2,

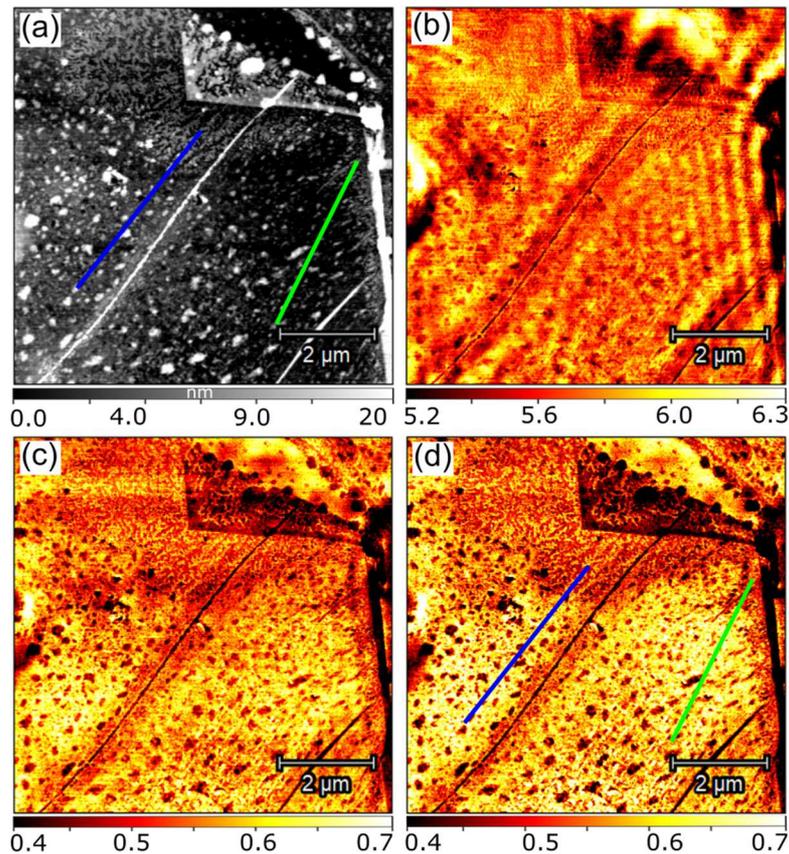

Figure 2: (a) AFM topography (colour bar in nm) together with (b)-(d) s-SNOM images (colour bar arb. units) at the same sample area as Fig. 1. The green and blue lines represent line-scans in Fig. 3. (b)-(d) s-SNOM amplitude images demodulated at (b) the second, (c) the third, and (d) the fourth harmonic with 594 nm excitation.

Supplementary Information). To confirm these findings, TEPL measurements were conducted at specific positions of the sample as indicated in the PL map, TEPL spectra are shown in Figure 1(d). Like the conventional PL, TEPL spectra show partially quenched PL intensity in the adhered area. The PL intensity increases towards the non-adhered area of the WS$_2$ while exhibiting significant peak shifts (tip up/tip down TEPL spectra shown in Figure S3, Supplementary Information).

A possible reason for the darker KPFM contrast (lower surface potential) in the non-adhered area is carbonaceous contamination of the Au surface which prevents charge transfer. Mechanical exfoliation of sulphur based TMDC's on Au substrates is facilitated by Au's affinity for sulphur, which is stronger than the van der Waals forces between the layers of the bulk TMD crystal [33]. With a prolonged (minutes) exposition of the Au surface to air [33], [34], this can lead to the accumulation of airborne organic contaminants on the Au surface, creating a buffer between the Au substrate and the WS$_2$ sample. This very likely applies here as the mechanical exfoliation was done on an Au substrate by PDMS transfer, which due to the nature of viscoelasticity requires a slow and steady exfoliation process [21]. This is further supported by the unquenched PL area as it suggests that the charge transfer in this area has been inhibited. In addition, the variations of the PL position and intensity in this area indicates different levels of local strain and doping throughout the WS$_2$ layer caused by the transfer method [8].

The surface roughness and sample contaminations lead to local variations in the dielectric function of WS$_2$, which manifests as a dielectric disorder [10]. It is also important to note that the dielectric function of WS$_2$ has a resonance peak around 620 nm (1.98 eV) [35].

Stress and strain indicated by the PL peak shifts could therefore cause a strong shift in the dielectric value [36].

The dual s-SNOM offers the opportunity to record optical images with nanometre resolution, and from the near field contrast variation to recover the dielectric values at different frequencies and at different fundamental harmonics $n$ of the tip frequency. The lower the harmonic, the higher the bulk sensitivity [25]. We recorded s-SNOM images of the pre-characterized area at excitation wavelengths of 594 nm, 604 nm and 614 nm (Figure 2 and Figure S4, Supporting Information). Similar to the KPFM measurements, s-SNOM provides an AFM scan along with the near field scans which is shown in Figure 2(a). All the previously identified features can be identified in this AFM scan. Figure 2(b) shows an s-SNOM image of the sample area demodulated at the second harmonic ($n = 2$). The fringes seen in this image are attributed to surface plasmon polaritons (SPPs) from the Au substrate underneath the WS$_2$ with a propagation constant of $k_{\parallel} \approx 0.9 \times 10^{-2}\ nm^{-1}$ which is similar to Takagi et al.[37], that observed SPP's in Au at this excitation with a propagation constant of $k_{\parallel} \approx 1.1 \times 10^{-2}\ nm^{-1}$, also observed in other previous studies [38], [39], [40], illustrating the subsurface sensitivity of the second harmonic. Interestingly, the characteristic areas identified in the KPFM, and PL measurements are not observable in the second harmonic s-SNOM image. Figure 2(c), the third harmonic ($n = 3$), and (d), the fourth harmonic ($n = 4$), clearly show that the higher the harmonic, the easier it is to recognize the various features and correlate them to areas identified in Figure 1; the triangular bilayer and the area surrounding it have a darker contrast than the detached WS$_2$ area, clearly correlating with the quenching of PL and the brighter KPFM contrast.

Thus, to study these contrasting areas of the monolayer WS$_2$, we consider the more surface sensitive fourth harmonic s-SNOM image. As previously mentioned, the near field contrast is sensitive to the distance between the tip and sample, and to the local dielectric value [41]. In comparing the s-SNOM images (Figure 2 (b), (c), (d)) with the AFM, one can see the effect of bubbles and wrinkles on the near field contrast. The long diagonal wrinkle starting from the bilayer and moving to the bottom left appears in the near field scan as a long black line due to the large height separation between the tip and the Au substrate over this wrinkle. This can be seen in all the bubbles and in another wrinkle on the right-hand side of the s-SNOM images. There is however only negligible height variation between the dark and the bright near field contrast areas in comparison to the wrinkles. The change in contrast between these two areas therefore must stem from variations in the local dielectric value. Using the inversion method (see methods), we were able to transform near field line-scans into line-scans featuring the local dielectric constant from the sample. These results are shown in Figure 3.

Figure 3 (a) and (d) show line-scans of the near field contrast $\eta_4$, which were extracted along the green and blue lines, respectively, featured in Figure 2. These line-scans were chosen as they cross the border of the areas with dark and bright near field contrast. Both line-scans exhibit a step in near field contrast from dark (~0.45) to bright (~0.65) contrast. Evaluating these line-scans through the inversion method provides the local dielectric constant of the sample (Figure 3(b) and (e)). The results suggest that the darker contrast area, corresponding to the area of strong WS$_2$/Au adhesion, has a dielectric value (at 594nm) that fluctuates around $-9$, while the brighter surrounding area has a less negative dielectric value of around $-6$ (both values are the real part of the dielectric function). The dielectric value increases with a sharp step between the two areas, which is highlighted in grey. Figure 3(c) and (f) feature the AFM topography of these line-scans, to analyse the effect of the tip height from the sample on the near field contrast. However, the height changes do not show any correlation to the step seen in the near field scans, both of which were measured at the same time by the same tip.

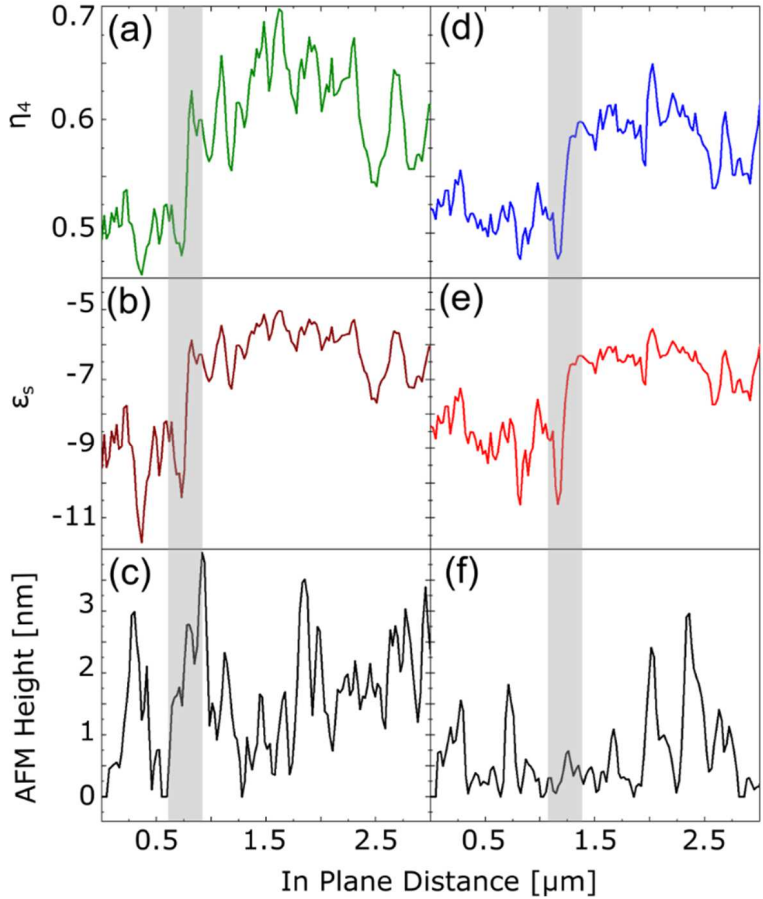

Figure 3: (a) s-SNOM near field contrast of the fourth harmonic, recorded along the colour coded lines indicated in Figure 2 (a). (b), (e) Extracted dielectric values using the inversion method on (a) and (d). A step in the real dielectric value can be seen from -9 to -6, highlighted in grey. (c), (f) AFM linescans along the same line-scan as (a) and (d) showing a topography that on average does not vary enough to produce the sharp step in contrast in (a) and (f).

To confirm the results of the inversion method shown in Figure 3, we performed SIE on the sample which is a well-established far-field method for measuring the dielectric function [42]. SIE measurements were performed on the dark and bright near field contrast area with the results depicted as solid black and solid red lines in Figure 4(b) and (c). SIE measurements carried out on the Au substrate are shown in Figure 4(b) and (c) as a solid yellow line. The spectral dependence of the dielectric function of free standing $WS_2$ is shown in Figure 4(b) and (c) as a dashed grey line [35], offset by a negative value to account for the influence of the Au substrate. SIE has a lateral resolution of approximately 1 µm. For comparison to s-SNOM, we extracted a 1 µm-by-1 µm area from the s-SNOM images (Figure S4, Supporting Information) and calculated the average near field contrast for $n = 2$ and $n = 4$. Figure 4(b) and (c) show the average dielectric values for $n = 2$ and $n = 4$ respectively at different excitation wavelengths. The black and red data points are the dielectric values from the bright and dark near field contrast areas (see Figure S4, Supplementary Information for SNOM images and areas for averaging). Figure 4(b) shows good agreement between the extracted dielectric values from $n = 2$ s-SNOM contrast and the ellipsometry measurements. In contrast, the dielectric values from the $n = 4$ s-SNOM contrast deviate from the ellipsometry measurements (Figure 4(c)). This is expected as ellipsometry is a far field technique that interacts considerably more with the bulk. The penetration depth of SIE through a metal surface is roughly 25 nm at 625 nm [43]. Considering the monolayer thickness of the WS$_2$, the penetration depth of SIE in this case makes it a good match for the second harmonic. The fourth harmonic ($n = 4$) data points, featured in Figure 4(c), are more surface sensitive and thus do not follow the same trend as the ellipsometry measurements. The determined dielectric values are in the same negative range (between -6 and -9) as the second harmonic but they exhibit a slope with a positive trend, emulating free standing WS$_2$. The black data points diverge more strongly from bulk Au than the red data points. This shows that WS$_2$ modifies the dielectric function of Au more in the adhered area than in the less adhered area. Due to the surface sensitivity of the fourth harmonic, we can resolve this dielectric modification from the monolayer WS$_2$ which is not visible in the second harmonic trend or in the SIE data. The larger sensitivity of the fourth harmonic to the surface dielectric disorder is also documented by the larger variation in near field contrast, expressed by the error bars in Figure 4(c) compared with the negligible error bars in Figure 4(b) for the second harmonics.

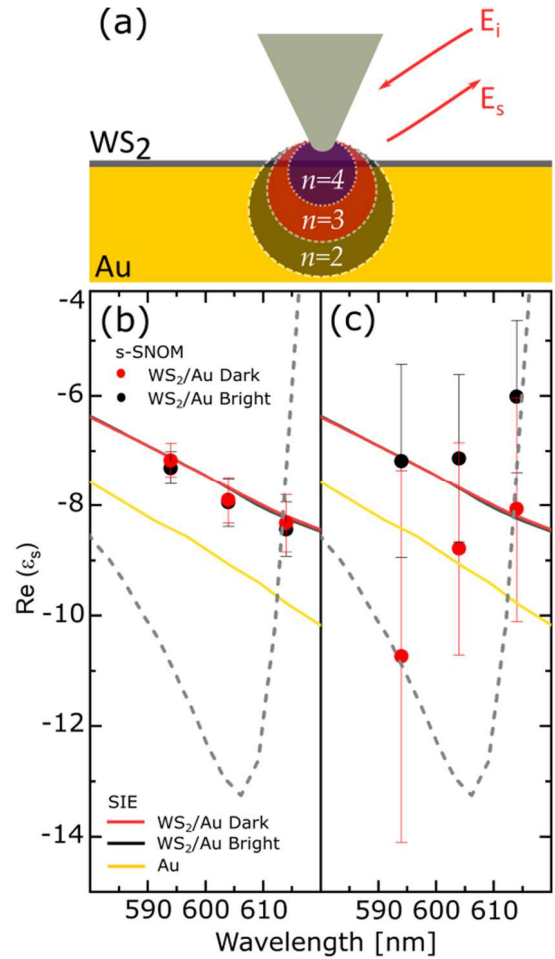

Figure 4: (a) Sketch of the SNOM tip above the Au/WS$_2$ sample. The blue, red, and black transparent circles show the penetration depth of each harmonic. The exact difference in depth between harmonics is difficult to determine but pervious works quote the optical near field penetrating SiO$_2$ as far as 100 nm, for Au we expect less penetration [23]. (b) Real part of the dielectric value as a function of wavelength from SIE for the darker contrast area (solid red line) and brighter contrast area (solid black line), a bare Au surface (yellow solid line). The dashed grey line is free standing WS$_2$ offset by a negative value to account for the influence of the Au substrate. Data points show the corresponding extracted values from s-SNOM by demodulating the second harmonic (n=2). Error bars show the standard deviation of the average value from the sample area. (c) same as (b) but the s-SNOM data used for the inversion method were taken from the fourth harmonic (n=4).

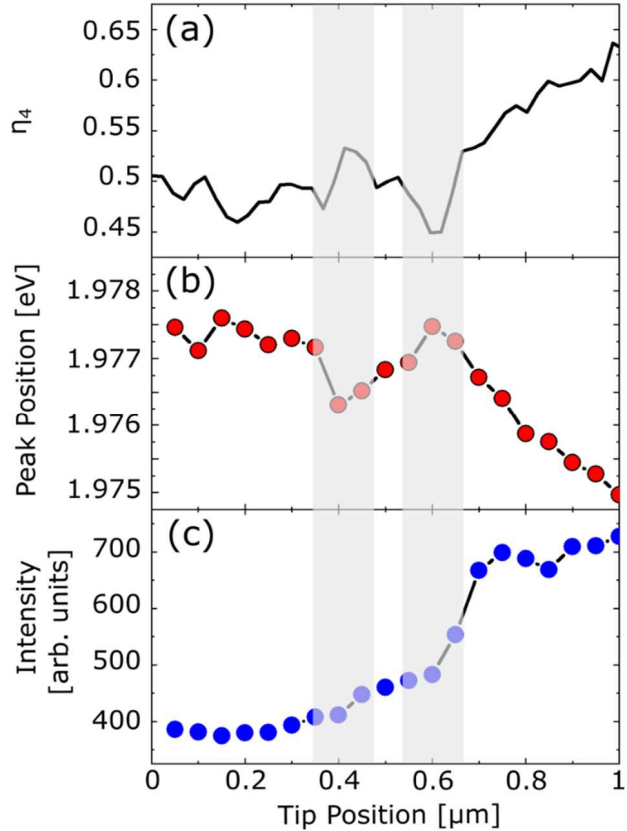

Figure 5: (a) Fourth harmonic (n=4) s-SNOM line-scan. (b) shows the TEPL peak position as a function of tip position taken at 50 nm. The sample data was taken from a 1 μm area at the near field intensity step. (c) TEPL intensity variation from the dark near field area to the bright near field area. The grey stripes highlight two features in the near field and TEPL peak position that have an inverse correlation. This highlights the sensitivity of the dual s-SNOM, in this case of 50 nm.

To visualise the influence of dielectric disorder at the nanometre scale we performed TEPL measurements (Figure 5). We recorded PL spectra as a function of tip position along a line moving from the dark to the bright contrast area in steps of 50 nm (line across the blue, red, and black dots in Fig. 1(e), raw spectra in Figure S6, Supporting Information). We extracted the peak position from the TEPL data (Figure 5 (b)) and compared it with the near field contrast of the fourth harmonic (Figure 5(a)). Figure 5(c) shows the PL intensity as a function of tip position where one can clearly see the transition from the dark contrast area to the bright. The grey stripes highlight two features in the near field contrast and the TEPL peak position where there is an increase in the former and a decrease in the latter. It is also clear that where the near field contrast increases, the peak position decreases. This demonstrates a resolution of 50 nanometres for both near field contrast variations and TEPL peak position changes. Additionally, since the near field contrast can be used to calculate the dielectric function of the sample, this leads to local dielectric resolution on the same scale as the near field signal.

With the ability to resolve the local dielectric environment at the nanometre scale, in addition to having access to sub-surface information as shown by Figure 4, the dual s-SNOM system is an excellent choice for characterising dielectric disorder. Since the monolayer surface is not uniformly flat, this leads to a varying Coulombic interaction between charge carriers resulting in local fluctuations of the permittivity [10]. This in turn produces spatial inhomogeneities of exciton binding energies, which can be seen from the PL peak position shifts in the bright near field contrast area of the sample (Figure 5(b) and Figure S6). KPFM and PL mapping (Figure 1) showed two areas with different levels of charge transfer happening between sample and substrate. With TEPL measurements recorded by the dual s-SNOM featured in Figure 5 we were able to come to the same conclusion at much higher resolution, evidenced by a defined border and sample roughness, in the fourth harmonic s-SNOM image of Figure 2(d).

Considering the capability of the s-SNOM shown in this work, it can be used for a variety of applications. Using different TMDC samples will provide deeper insights into the sample-substrate interaction. The sub-surface sensitivity of the fourth harmonic also enables the study of sandwiched 2D samples such as heterostructures encased in an insulator, like $WSe_2/MoS_2$ encapsulated in hBN for example. Additionally, because the insights into the dielectric constant it provides, as well as the resolution, it could be useful in experimentally determining the dielectric function of low dimensionality systems, such as individual CNTs and graphene nanoribbons.

## 5. Conclusion

We demonstrated the capability of the dual s-SNOM in measuring the spatially resolved dielectric values of nanoscale systems at different excitation energies. This enables the determination of the dielectric function with a resolution on the nanometre scale. As an example, we used a monolayer sample of $WS_2$ exfoliated on Au. By comparing the s-SNOM characterisation of this sample with conventional sample characterisation methods like KPFM and SIE, and far field techniques like conventional PL mapping, we illustrated the superior resolution of the dual s-SNOM in comparison to these techniques. The dual-SNOM also provided local dielectric information as a result of the near field contrast being sensitive to the local dielectric environment of the sample. This was illustrated using an inversion method with which we extracted the local dielectric values at different wavelengths and, thanks to the selective penetration depths of the different near field image harmonics, from the sub-surface and surface of the sample. We believe this could be useful for identifying and characterising interlayer excitons by probing dielectric differences in the sample environment, probing sandwiched TMDC heterostructures using the different harmonics penetration depth, and determining the dielectric function of low-dimensional systems like carbon nanotubes and graphene nanoribbons.

## Acknowledgments

A.R. acknowledges financial support of the Czech Science Foundation (project no. 20-08633X), O.F. appreciates the support by European Regional Development Fund; OP RDE; Project: "Carbon allotropes with rationalized nanointerfaces and nanolinks for environmental and biomedical applications" (No. CZ.02.1.01/0.0/0.0/16_026/0008382).

N.S.M. acknowledges support from the German National Academy of Sciences Leopoldina.

P.K. acknowledge the German Science Foundation (DFG) within the Priority Program SPP 2244 2DMP for funding.

# Supplement: Probing the Local Dielectric Function by Near Field Optical Microscopy Operating in the Visible Spectral Range

Authors:

Oisín Garrity[a], Alvaro Rodriguez[b], Niclas S. Mueller[c], Otakar Frank[b], Patryk Kusch[a]*

Affiliations:

a) Department of Physics, Freie Universität Berlin, Arnimallee 14, D-14195 Berlin, Germany
b) J Heyrovský Institute of Physical Chemistry, Academy of Sciences of the Czech Republic, Dolejškova 3, CZ-18223 Prague 8, Czechia
c) NanoPhotonics Centre, University of Cambridge, UK

*corresponding author: patryk.kusch@fu-berlin.de

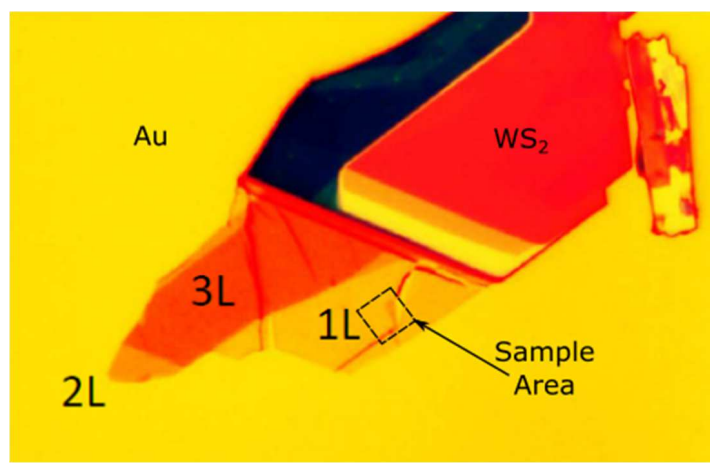

*Figure S1: Optical microscope image of the sample. Black square indicates the sample area of this work, within which a triangular piece of bilayer can be seen.*

Figure S1 shows an optical microscope image of the sample used in this work. The black square indicates the sample area that was studied, looking closely you can see a triangular piece of bilayer which is highlighed magenta in Figure 1(a).

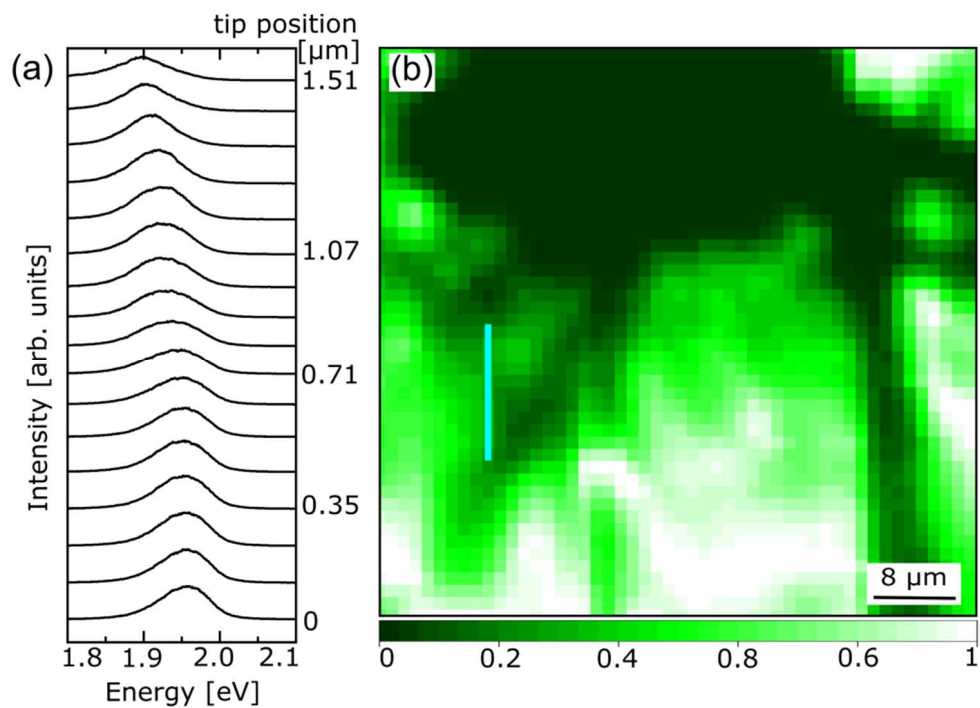

Figure S2: (a) PL spectra as a function of tip position, (b) PL intensity map with blue line indicating the location of the spectra in (a).

Figure S2(a) shows PL spectra as a function of tip position extracted from the blue line seen in (b) the PL map. This shows an example of the peak shifts common throughout the sample.

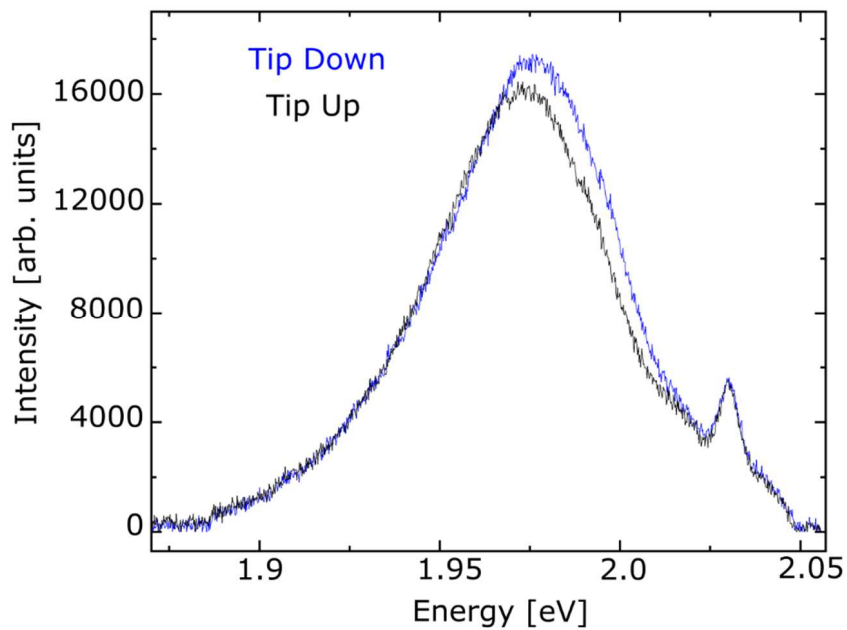

Figure S3: TEPL spectra showing the effect of tip up (black) and tip down (blue). Tip down shows influence of the dark area with a shift to higher energy while tip up shows less intensity from the dominant conventional PL from the bright area with a peak shift to lower energies.

    Figure S3 shows tip enhanced photoluminescence (TEPL) spectra that were measured close to the transition from dark to bright near field contrast, blue for when the tip was down, and black from when the tip was up. Once the tip is down, you can see slightly larger intensity and a shift of the peak, (dark area means shift to higher energy, shown in Figure 5). Once the tip is up, the intensity decreases slightly as the conventional PL from the bright area dominates and we see the PL peak position shift to lower energy.

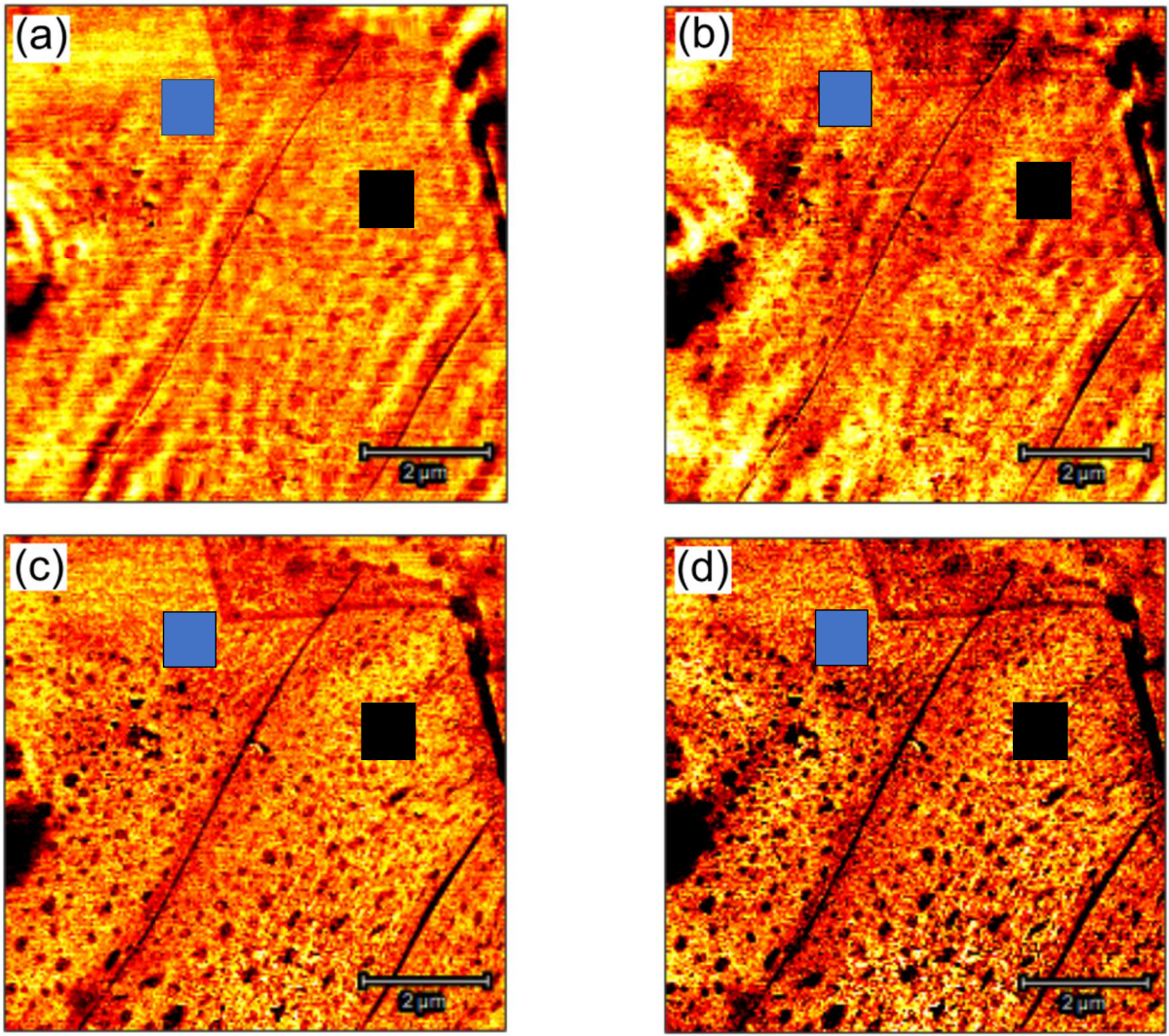

*Figure S4: s-SNOM amplitude images, top row second harmonic for (a) 604 nm and (b) 614 nm, bottom row fourth harmonic for (c) 604 nm and (d) for 614 nm. Red and black squares show where the values were taken for the average to compare to ellipsometry.*

The s-SNOM images used for the ellipsometry comparison in Figure 4 of the paper where 1μm x 1 μm squares of near field contrast values were taken from the bright area (black square) and the dark area (red square). The top row in Figure S4 show the second harmonic measurements, the bottom row shows the fourth harmonic measurements; (a) and (c) at 604 nm and (b) and (d) at 614 nm.

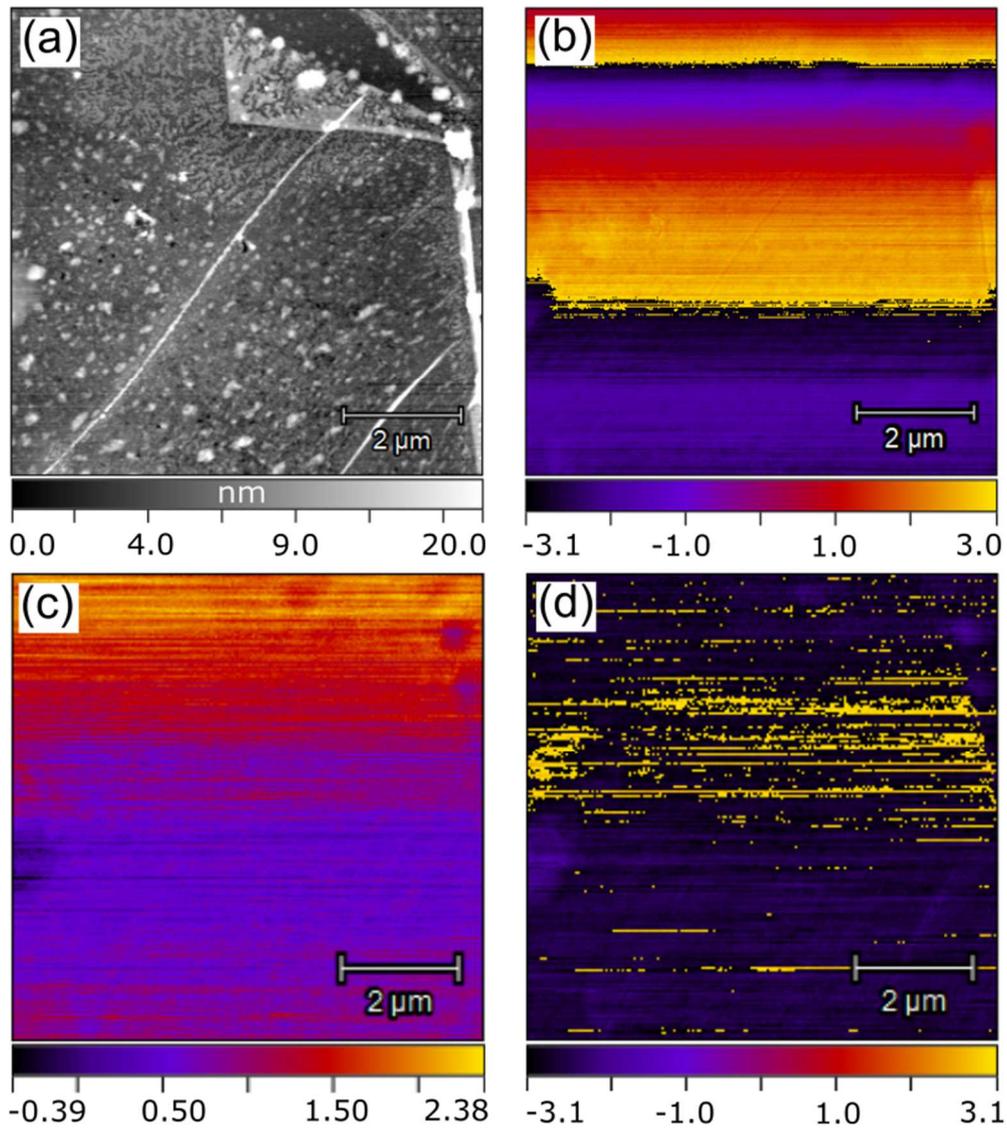

*Figure S5: (a) AFM image taken along with (b) n=4 s-SNOM phase image with excitation of 594nm. (c), (d) n=4 s-SNOM phase images taken with excitation of 604 nm and 614 nm respectively.*

The phase images that accompany the amplitude measurements seen in Figure 2 are seen in Figure S5 with the AFM scan included for the sake of comparison. As a result of the poor quality of the phase images, the imaginary part of the dielectric function could not be reliably extracted.

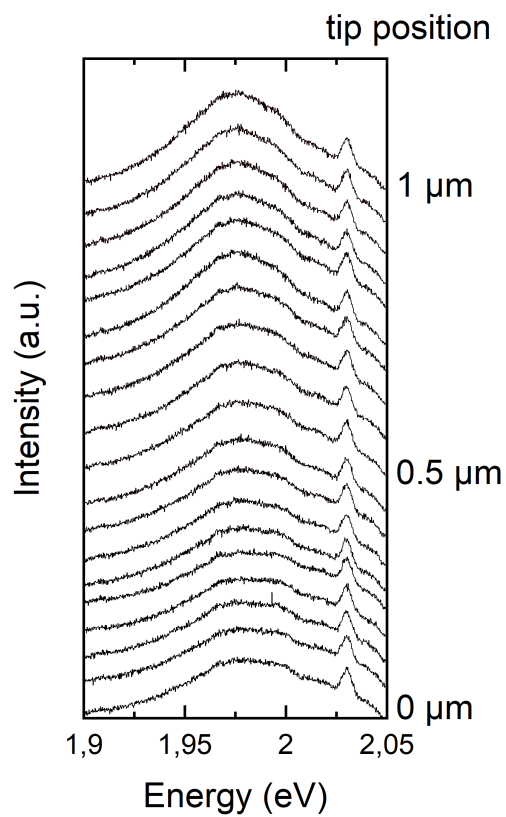

*Figure S6: TEPL spectra as a function of tip position. The PL position is plotted as function of tip position in Fig. 5*

    In Fig. S6 we show the tip-enhanced photoluminescence spectra recorded as a function of tip position. From the spectra we estimate the PL position and plot it as a function of tip position shown in Fig. 5 in the main manuscript.